\documentclass[12pt]{article}
\usepackage{jheppub,subcaption}

\title{Landau Singularities from the Amplituhedron}

\author[1]{T.~Dennen%
\note{Current address: Google Inc., Mountain View CA 94043, USA},}
\author{I.~Prlina,}
\author{M.~Spradlin,}
\author{S.~Stanojevic}
\author{and A.~Volovich}

\affiliation{Department of Physics, Brown University,
Providence RI 02912, USA}

\abstract{We propose a simple geometric algorithm for
determining the complete set of branch points of amplitudes
in planar $\mathcal{N} = 4$ super-Yang-Mills theory directly from
the amplituhedron, without resorting to any particular representation
in terms of local Feynman integrals. This represents
a step towards translating integrands
directly into integrals. In particular,
the algorithm provides information about the
symbol alphabets of general amplitudes. We illustrate the algorithm
applied to the one- and two-loop MHV amplitudes.
}

\begin{document}
\maketitle

\section{Introduction}
\label{sec:introduction}

Ever since its conception, the Feynman diagram approach has been the standard paradigm for perturbative calculations in quantum field theory.
While the method can, in principle, be used at any order in perturbation theory, the calculations get more and more demanding at each new loop order.
Alternately one can seek hidden symmetries and new underlying principles which motivate new calculational approaches where the most basic features of Feynman diagrams, such as unitarity and locality, are emergent instead of manifest.
Recent years have seen tremendous success in ``reverse engineering''
such
new symmetries and principles from properties of scattering amplitudes.
This approach has been particularly fruitful in simple quantum field
theories such as the planar maximally supersymmetric
$\mathcal{N} = 4$ super-Yang-Mills (SYM) theory~\cite{Brink:1976bc}.

In particular, it has been realized
that the unitarity and locality of the integrands~\cite{ArkaniHamed:2010kv}
of loop-level
amplitudes in SYM
theory can be seen to emerge from a very simple
geometric principle of positivity~\cite{ArkaniHamed:2012nw}.  Moreover,
it has been proposed that all information about arbitrary
integrands in this theory is encapsulated in objects called
amplituhedra~\cite{Arkani-Hamed:2013jha,Arkani-Hamed:2013kca} that
have received considerable recent attention; see for
example~\cite{Bai:2014cna,Franco:2014csa,Lam:2014jda,Arkani-Hamed:2014dca,Bai:2015qoa,Ferro:2015grk,Bern:2015ple,Galloni:2016iuj}.
Unfortunately, there remains a huge gap between our understanding
of integrands and our understanding of the corresponding
integrated amplitudes.
Despite great advances in recent years
we of course  don't have a magic wand that can be waved at
a general integrand to ``do the
integrals''.
Indeed, modern approaches to computing multi-loop
amplitudes in SYM theory, such as the amplitude
bootstrap~\cite{Dixon:2014xca,Golden:2014pua}
even eschew knowledge of the integrand completely.
It would be enormously valuable to close this gap between our
understanding of integrands
and amplitudes.

As a step in that direction, and motivated by~\cite{Maldacena:2015iua},
we began
in~\cite{Dennen:2015bet}
to systematically explore how integrands encode
the singularities
of integrated amplitudes, in particular their branch points.
Scattering amplitudes in quantum field theory generally
have very complicated discontinuity structure.  The discontinuities
across branch cuts are given by sums of
unitarity cuts~\cite{Mandelstam:1958xc,Mandelstam:1959bc,Landau:1959fi,Cutkosky:1960sp,Bern:1994cg,Bern:1996je}.
These discontinuities may appear on the physical sheet or after
analytic continuation to other sheets; these higher discontinuities
are captured by multiple unitarity cuts (see
for example~\cite{Abreu:2014cla,Abreu:2015zaa}).
A long-standing goal of the S-matrix program, in both its
original and modern incarnations, has been to construct expressions
for the scattering amplitudes of a quantum field theory
based solely only on a few physical principles and a thorough knowledge
of their analytic structure.

In~\cite{Dennen:2015bet} we studied the branch cut
structure of one- and two-loop
MHV amplitudes in SYM theory starting from certain
representations of their integrands in terms of local Feynman
integrals~\cite{ArkaniHamed:2010gh}.
We recovered all of their known branch points,
but we also encountered many other, spurious branch points
that are artifacts of the particular representations used.
Indeed, the analysis of~\cite{Dennen:2015bet} was completely insensitive
to numerator factors in the integrand, but the numerators are really
where all of the action is---in any standard quantum field theory the
denominator of a loop integrand is a product of local propagators;
the numerator is where all of the magic lies.

Our goal in this paper is to improve greatly on the analysis
of~\cite{Dennen:2015bet}. We do this by presenting a method for asking
the amplituhedron to directly provide a list of the
physical branch points of a given amplitude.
In the remainder of section 1 we briefly review the necessary
background on momentum twistor notation, the MHV amplituhedron,
and Landau singularities.
In section 2 we demonstrate how to refine the analysis
of~\cite{Dennen:2015bet} by scanning through the list of putative
branch points found in that paper, and asking the amplituhedron
to identify each one as physical or spurious.
This is an ultimately inefficient approach, but armed with
experience from that exercise we turn in section 3 to the
development of a general, geometric algorithm for reading
off the physical branch points of MHV amplitudes directly
from the amplituhedron.

\subsection{Momentum Twistors}

We begin by reviewing the basics of momentum twistor notation~\cite{Hodges:2009hk}, which we use throughout our calculations. Momentum twistors are based on the correspondence between null rays in (complexified, compactified) Minkowski space and points in twistor space ($\mathbb{P}^3$), or equivalently, between complex lines in $\mathbb{P}^3$ and points in Minkowski space. We use $Z_a$, $Z_b$, etc.~to denote points in $\mathbb{P}^3$, which may be represented using four-component homogeneous coordinates $Z^I_a = (Z_a^1, Z_a^2, Z_a^3, Z_a^4)$ subject to the identification $Z^I_a \sim t Z^I_a$ for any non-zero complex number $t$. We use $(a \, b)$
as shorthand for the bitwistor $\epsilon_{IJKL} Z_a^K Z_b^L$.
Geometrically, we can think of $(a \, b)$ as the (oriented) line containing the points $Z_a$ and $Z_b$.  Similarly we use $(a \, b \, c)$ as shorthand
for $\epsilon_{IJKL} Z_a^J Z_b^K Z_c^L$, which represents the (oriented) plane
containing $Z_a$, $Z_b$ and $Z_c$.
Analogously, $(a \, b \, c) \cap (d \, e \, f)$ stands for $\epsilon^{IJKL} (a \, b \, c)_K (d \, e \, f)_L$, which represents the line where the two indicated planes intersect. In planar SYM theory we always focus on color-ordered partial amplitudes so an $n$-point amplitude is characterized by a set of $n$ momentum twistors $Z^I_i$, $i \in \{ 1, \ldots, n \}$ with a specified cyclic ordering. Thanks to this implicit cyclic ordering we can use $\bar{i}$ as shorthand for the plane $(i{-}1\,i\,i{+}1)$, where indices are always understood to be mod $n$.

The natural $SL(4,\mathbb{C})$ invariant is the four-bracket
denoted by
\begin{equation}
\langle a \, b \, c \, d \rangle \equiv \epsilon_{I \, J \, K \, L} Z_a^I Z_b^J Z_c^K Z_d^L\,.
\end{equation}
We will often be interested in a geometric understanding of the
locus where
such four-brackets might vanish, which can be pictured in several
ways.
For example, $\langle a\,b\,c\,d\rangle = 0$
only if the two lines $(a \, b)$ and $(c \, d)$ intersect, or equivalently if
the lines $(a \, c)$ and $(b \, d)$ intersect,
or if the point $a$ lies in the plane $(b \, c \, d)$, or if the point $c$ lies
on the plane $(a \, b \, d)$, etc.
Computations of four-brackets involving intersections may be simplified
via the formula
\begin{equation}
\langle (a \, b \, c) \cap (d \, e \, f) \, g \, h\rangle
= \langle a\, b\, c\, g \rangle \langle d\, e\, f\, h\rangle
- \langle a\, b\, c\, h \rangle \langle d\, e\, f\, g\rangle\,.
\end{equation}
In case the two planes are specified with one common point,
say $f=c$, it is convenient to use the shorthand notation
\begin{equation}
\langle (a \, b \, c) \cap (d \, e \, c)\, g\, h\rangle \equiv
\langle c\, (a \, b) (d \, e) (g \, h) \rangle
\end{equation}
which highlights the fact that this quantity is antisymmetric
under exchange of any two of the three lines $(a \, b)$, $(d \, e)$, and
$(g \, h)$.

\subsection{Positivity and the MHV Amplituhedron}

In this paper we focus exclusively on MHV amplitudes.
The integrand of an $L$-loop MHV amplitude is a rational function of the $n$ momentum twistors $Z_i$ specifying the kinematics of the $n$ external particles, as well as of $L$ loop momenta, each of which corresponds to some line $\mathcal{L}^{(\ell)}$ in $\mathbb{P}^3$; $\ell \in \{1, \ldots, L\}$.
The amplituhedron~\cite{Arkani-Hamed:2013jha,Arkani-Hamed:2013kca}
purports to provide a simple characterization of the
integrand when the $Z^I_i$ take values in a particular domain called the
positive Grassmannian $G_+(4, n)$. In general $G_+(k, n)$ may be defined
as the set of $k \times n$ matrices for which all ordered maximal
minors are positive; that is, $\langle a_{i_1} \cdots a_{i_k} \rangle > 0$ whenever
$i_1 < \cdots < i_k$.

Each line $\mathcal{L}^{(\ell)}$ may be characterized by specifying
a pair of points $\mathcal{L}^{(\ell)}_1$,
$\mathcal{L}^{(\ell)}_2$ that it passes through.
We are always interested in $n \ge 4$, so the $Z_i$ generically
provide a basis for $\mathbb{C}^4$.
In the MHV amplituhedron a pair
of points specifying each $\mathcal{L}^{(\ell)}$
may be expressed in the $Z_i$ basis via
an element of $G_+(2, n)$ called the $D$-matrix:
\begin{equation}
\label{eq:dmatrixdef}
\mathcal{L}_\alpha^{(\ell)I} = \sum_{i=1}^n D_{\alpha i}^{(\ell)} Z_i^I, \qquad
\alpha = 1, 2\,.
\end{equation}
For $n > 4$ the $Z_i$ are generically overcomplete, so
the map eq.~(\ref{eq:dmatrixdef}) is many-to-one.

The $L$-loop $n$-point MHV amplituhedron is a $4L$-dimensional
subspace of the $2L(n-2)$-dimensional space of $L$ $D$-matrices.
We will not need a precise characterization of that subspace, but only
its grossest feature, which is that
it is a subspace of the space of $L$ mutually
positive points in $G_+(2,n)$.
This means that it lives in the subspace for which all ordered maximal
minors of the matrices
\begin{equation}
\begin{pmatrix}
        D^{(\ell)}
        \end{pmatrix},
\quad
\begin{pmatrix}
        D^{(\ell_1)} \\
        D^{(\ell_2)}
        \end{pmatrix},
\quad
\begin{pmatrix}
        D^{(\ell_1)} \\
        D^{(\ell_2)} \\
	D^{(\ell_3)} \\
        \end{pmatrix},
\quad \text{etc.}
\nonumber
\end{equation}
are positive.

A key consequence of the positivity of the $D$-matrices is that, for positive external data $Z^I_i\in G_+(4,n)$, all loop variables $\mathcal{L}^{(\ell)}$ are oriented positively with respect to the external data and to each other: inside the amplituhedron,
\begin{align}
\label{eq:inside1}
\langle \mathcal{L}^{(\ell)}\, i\, i{+}1 \rangle &> 0 \text{ for all } i \text{ and all } \ell, \text{ and} \\
\langle \mathcal{L}^{(\ell_1)}\, \mathcal{L}^{(\ell_2)} \rangle &> 0 \text{ for all } \ell_1, \ell_2.
\label{eq:inside2}
\end{align}
The boundaries of the amplituhedron coincide with the boundaries of the space of positive $D$-matrices, and occur for generic $Z$ when one or more of these quantities approach zero.

It is worth noting that the above definition of positivity depends
on the arbitrary choice of a special point $Z_1$, since for example
$\langle \mathcal{L}\, 1 \, 2\rangle > 0$ but
the cyclically related quantity
$\langle \mathcal{L}\, n \, 1 \rangle$ is negative.
The choice of special point is essentially irrelevant: it just means
that some special cases need to be checked.
In calculations we can sidestep this subtlety by
always choosing to analyze
configurations involving
points satisfying $1 \le i < j < k < l \le n$, which
can be done without loss of generality.
The geometric properties of figures~\ref{fig2}--\ref{fig5}
below are insensitive to the choice and always have full cyclic
symmetry.

The integrand of an MHV amplitude is a canonical form $d \Omega$ defined by
its having logarithmic singularities only on the boundary of the amplituhedron.
The numerator of $d \Omega$ conspires to cancel all singularities that
would occur outside this region (see~\cite{Arkani-Hamed:2014dca} for some detailed examples).
Our analysis will require no detailed knowledge of this form.
Instead, we will appeal to ``the amplituhedron'' to tell
us whether or not any
given configuration of lines $\mathcal{L}^{(\ell)}$
overlaps the amplituhedron or its boundaries by checking whether
eqs.~(\ref{eq:inside1}) and~(\ref{eq:inside2}) are satisfied (possibly
with some $=$ instead of $>$).

\subsection{Landau Singularities}

The goal of this paper is to understand the singularities of (integrated)
amplitudes.  For standard Feynman integrals, which are characterized
by having only local propagators in the denominator, it is well-known
that the locus in kinematic space where a Feynman integral can potentially
develop a singularity is determined by solving the
Landau equations~\cite{Landau:1959fi,Coleman:1965xm,ELOP}
which
we now briefly review.

After Feynman parameterization any $L$-loop
scattering amplitude in $D$ spacetime dimensions may be expressed
as a linear combination of integrals of the form
\begin{equation}
\int \prod_{r=1}^L d^D l_r \int_{\alpha_i \geq 0} d^\nu \alpha\ \delta \left(   1-\sum_{i=1}^\nu \alpha_i   \right) \frac{{\cal{N}}(l_r^\mu, p_i^\mu,...)}{{\cal{D}}^\nu}
\end{equation}
where $\nu$ is the number of propagators in the diagram, each of which
has an associated Feynman parameter $\alpha_i$,
$\mathcal{N}$ is some numerator factor which may depend
on the $L$ loop momenta $l^\mu_r$ as well as the external momenta
$p_i^\mu$, and finally the denominator involves
\begin{equation}
\mathcal{D}=\sum_{i=1}^\nu \alpha_i (q_i^2-m_i^2)\,,
\end{equation}
where $q_i^\mu$ is the momentum flowing along propagator $i$ which carries
mass $m_i$.
The integral can be viewed as a multidimensional contour integral
in the $L D + \nu$ integration variables
$(l_r^\mu, \alpha_i)$, where the $\alpha_i$ contours begin at $\alpha_i = 0$
and the $l_r^\mu$ contours are considered
closed by adding a point at infinity.
Although the correct contour for a physical scattering process
is dictated by an appropriate $i \epsilon$ prescription in the propagators,
a complete understanding
of the integral, including its analytic continuation off the physical
sheet, requires arbitrary contours to be considered.

An integral of the above type can develop singularities when the denominator
$\mathcal{D}$ vanishes in such a way that the contour of integration
cannot be deformed to avoid the singularity.
This can happen in two distinct situations:

(1) The surface
$\mathcal{D}=0$ can pinch the contour simultaneously in
all integration variables $(l_r^\mu, \alpha_i)$.  This is called
the ``leading Landau singularity'', though it is important to keep
in mind that it is only a potential singularity.  The integral may
have a branch point instead of a singularity, or it may be a completely
regular point, depending on the behavior of the numerator factor
$\mathcal{N}$.

(2) The denominator may vanish on the boundary when one or
more of the $\alpha_i = 0$ and pinch the contour in the other integration variables.
These are called subleading Landau singularities.

The Landau conditions encapsulating both possible situations are
\begin{align}
\label{eq:LS1}
\sum_{i \in \text{loop}} \alpha_i q_i^\mu &= 0 \text{ for each loop, and} \\
\alpha_i (q_i^2 - m_i^2) &= 0 \text{ for each } i.
\label{eq:LS2}
\end{align}
For leading singularities eq.~(\ref{eq:LS2}) is satisfied by
$q_i^2 - m_i^2 = 0$ for each $i$, while subleading
singularities have one or more $i$ for which $q_i^2 - m_i^2 \ne 0$
but the corresponding
$\alpha_i = 0$.
We will always refer to equations of type $q_i^2 - m_i^2$ as ``cut
conditions'' since they correspond to putting
some internal propagators on-shell.
It is important to emphasize that the Landau equations themselves have
no knowledge of the numerator factor $\mathcal{N}$, which can alter
the structure of a singularity or even cancel a singularity entirely.

Sometimes (i.e., for some diagram topologies),
the Landau equations~(\ref{eq:LS1}) and~(\ref{eq:LS2}) may admit
solutions for arbitrary external kinematics $p_i^\mu$.
This usually indicates an infrared divergence in the integral (we will
not encounter ultraviolet divergences in SYM theory), which may or may
not be visible by integration along the physical contour.

In other cases,
solutions to the Landau equations might exist only when the
$p_i^\mu$ lie on some subspace
of the external kinematic space.
MHV amplitudes in SYM theory are expected to have only
branch point type singularities (after properly normalizing them
by dividing out a tree-level Parke-Taylor~\cite{Parke:1986gb} factor), so for
these amplitudes we are particularly
interested
in solutions which exist only on codimension-one slices of
the external kinematic space.
Even when the $p_i^\mu$ live on a slice where solutions of the
Landau equations exist, the solutions
generally occur
for values of the integration variables $\alpha_i$ and $l^\mu_r$
that are off the physical contour
(for example, the $\alpha_i$ could be complex).
This indicates a branch point of the integral that is not
present on the physical sheet but only becomes apparent after suitable
analytic continuation away from the physical contour.

Finally let us note that we have ignored a class of
branch points called
``second-type singularities''~\cite{Fairlie:1962-1,Fairlie:1962-2,ELOP}
which arise from pinch singularities at infinite loop momentum.
As argued in~\cite{Dennen:2015bet}, these should be absent
in planar SYM theory when one uses a regulator that preserves
dual conformal symmetry.

\section{Eliminating Spurious Singularities of MHV Amplitudes}
\label{sec:outline}

In principle one can write explicit formulas for any desired integrand
in planar SYM theory by
triangulating the interior of the amplituhedron and constructing the
canonical form $d \Omega$ with logarithmic singularities on its
boundary.
However, general triangulations may produce arbitrarily complicated
representations for $d \Omega$.  In particular, these may have no semblance
to standard Feynman integrals with only local propagators in
the denominator (see~\cite{Bai:2014cna} for some explicit examples).
It is therefore not immediately clear that the Landau equations have
any relevance to the amplituhedron.
The connection will become clear in the following section; here we begin
by revisiting the analysis of~\cite{Dennen:2015bet} with the amplituhedron
as a guide.

In~\cite{Dennen:2015bet} we analyzed the potential Landau singularities
of one- and two-loop MHV amplitudes by relying on the crutch of
representations of these amplitudes in terms of one- and
two-loop chiral pentagon and double-pentagon
integrals~\cite{ArkaniHamed:2010gh}.
The solutions to the various sets of Landau equations for
these integral topologies represent
{\it potential} singularities of the amplitudes, but this set
of potential singularities is too large for two reasons.
First of all, the chiral integrals
are dressed with very particular numerator factors to which the
Landau equations are completely insensitive.  Scalar pentagon and
double pentagon integrals certainly have singularities
that are eliminated by the numerator factors of their chiral
cousins.  Second, some actual singularities of individual chiral
integrals may be spurious in the full amplitude due to cancellations
when all of the contributing chiral integrals are summed.

It is a priori
highly non-trivial to see which singularities of individual
integrals survive the summation to remain singularities of
the full amplitude.  However, the amplituhedron hypothesis provides
a quick way to detect spurious singularities from simple
considerations of positive geometry.
In this section we refine our analysis of~\cite{Dennen:2015bet} to
determine which {\it potential} singularities identified in that paper are
{\it actual} singularities by appealing to the amplituhedron
as an oracle to tell us which cuts of the amplitude have zero or non-zero
support on the (boundary of the) amplituhedron.

Specifically, we propose
a check that is motivated by the Cutkosky rules~\cite{Cutkosky:1960sp},
which tell
us that to compute the cut of an amplitude with respect to some
set of cut conditions,
one replaces the on-shell propagators in the integrand
corresponding to those cut conditions by delta-functions,
and integrates the resulting quantity over the loop momenta.
The result of such a calculation has a chance to be non-zero only if the
locus where the cut conditions are all satisfied has non-trivial overlap
with the domain of integration of the loop momentum variables.
In the present context, that domain is the space of mutually positive
lines, i.e., the interior of the amplituhedron. This principle will lead to a 
fundamental asymmetry between the two types of Landau equations
in our analysis.
The full set of Landau equations
including both eqs.~(\ref{eq:LS1}) and~(\ref{eq:LS2})
should
be solvable only on a codimension-one locus in the space
of external momenta in order to obtain a valid branch point.
However, guided by Cutkosky, we claim that
the cut conditions~(\ref{eq:LS2}) must be solvable
inside the positive domain for arbitrary (positive) external kinematics; otherwise the discontinuity around the putative branch point is
zero and we should discard it as spurious.

In the remainder of
this section we will demonstrate this hypothesis by means of the
examples shown in figure~\ref{fig:1}.
The leading Landau singularities of each of these diagrams
were found to be singularities of the scalar pentagon and
double-pentagon integrals analyzed in~\cite{Dennen:2015bet},
but it is clear that MHV amplitudes have no support on these
cut configurations.  In the next three subsections we will
see how to understand their spuriousness directly from the amplituhedron.
This will motivate us to seek a better, more direct algorithm
to be presented in the following section.

\subsection{The Spurious Pentagon Singularity}

\begin{figure}[]
\centering
\subcaptionbox{\label{fig1:a}}{\raisebox{0.22cm}{\includegraphics[scale=1.44]{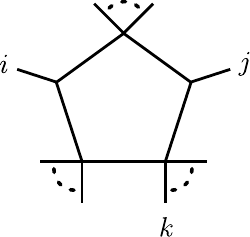}}}
\hfill
\subcaptionbox{\label{fig1:b}}{\raisebox{0.26cm}{\includegraphics[scale=1.44]{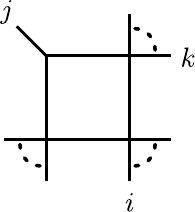}}}
\hfill
\subcaptionbox{\label{fig1:c}}{\includegraphics[scale=1.44]{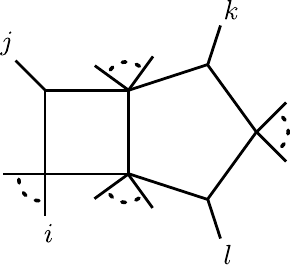}}
\caption{Three examples of cuts on which MHV amplitudes have no support;
these appeared as
spurious singularities
in the Landau equation
analysis of~\cite{Dennen:2015bet} since scalar pentagon and double
pentagon integrals do have these cuts.}
\label{fig:1}
\end{figure}

The first spurious singularity of MHV
amplitudes arising from the integral
representation used in~\cite{Dennen:2015bet} is the leading Landau
singularity of the pentagon shown in figure~\ref{fig1:a},
which is located on the locus where
\begin{equation}
\label{eq:oneloopbadLLS}
\langle i\, j\, k\, k{+}1 \rangle \langle \bar{i} \cap \bar{j}\ k\, k{+}1 \rangle = 0\,.
\end{equation}
It was noted already in~\cite{Cutkosky:1960sp}
that this solution of the Landau equations does not
correspond to a branch point
of the pentagon integral.
It arises from cut conditions that
put all five propagators of the pentagon on-shell:
\begin{equation}
\label{eq:oneloopbadLLScuts}
0 = \langle \mathcal{L} \, i{-}1\, i \rangle =
\langle \mathcal{L} \, i\, i{+}1 \rangle =
\langle \mathcal{L} \, j{-}1\, j \rangle =
\langle \mathcal{L} \, j\, j{+}1 \rangle =
\langle \mathcal{L} \, k\, k{+}1 \rangle\,,
\end{equation}
where $\mathcal{L}$ is the loop momentum.
The first four of these cut conditions admit two
discrete solutions~\cite{ArkaniHamed:2010gh}:
either $\mathcal{L} = (i\,j)$ or $\mathcal{L} = \bar{i} \cap \bar{j}$.
The second of these cannot avoid lying outside the
amplituhedron.
We see this by representing its $D$-matrix as
\begin{equation}
\label{eq:bad}
D =   \bordermatrix{ & i{-}1 & i & i{+}1 \cr
& \langle i\, \bar{j} \rangle & - \langle i{-}1\, \bar{j} \rangle & 0 \cr
& 0 & \langle i{+}1\, \bar{j} \rangle & - \langle i\, \bar{j}\rangle }\,,
\end{equation}
where we indicate only the
nonzero columns of the $2 \times n$ matrix
in positions $i{-}1$, $i$ and $i{+}1$,
per the labels above the matrix.  The non-zero $2 \times 2$ minors of this
matrix,
\begin{equation}
\langle i \, \bar{j} \rangle \langle i{+}1 \, \bar{j} \rangle,
\qquad \langle i{-}1\, \bar{j} \rangle \langle i\, \bar{j} \rangle,
\qquad - \langle i \, \bar{j} \rangle^2
\end{equation}
have indefinite signs for general positive external kinematics,
so this $\mathcal{L}$
lies discretely outside the amplituhedron.

We proceed with the first solution $\mathcal{L} = (i\, j)$ which
can be represented by the trivial $D$-matrix
\begin{equation}
D = \bordermatrix{ & i & j \cr
& 1 & 0 \cr
& 0 & 1 }\,.
\end{equation}
Although this is trivially positive, upon substituting
$\mathcal{L} = (i\, j)$ into eq.~(\ref{eq:oneloopbadLLScuts}) we find
that the fifth cut condition can only be satisfied for special
kinematics satisfying
\begin{equation}
\label{eq:oneloopbadLLSlocus}
\langle i\, j\, k\, k{+}1 \rangle = 0\,.
\end{equation}
Therefore, according to the Cutkosky-inspired rule discussed
three paragraphs ago, the monodromy
around this putative singularity vanishes for general kinematics
and hence it is not a valid branch point at one loop.
Indeed this conclusion is easily verified
by looking at the explicit results of~\cite{Bern:1994zx}.

\subsection{The Spurious Three-Mass Box Singularity}

The second spurious one-loop singularity encountered in~\cite{Dennen:2015bet} is a subleading singularity of the pentagon which lives on the locus
\begin{equation}
\label{eq:oneloopbadSLLS}
\langle j\, (j{-}1\, j{+}1) (i\, i{+}1) (k\, k{+}1) \rangle = 0
\end{equation}
and arises from the cut conditions
shown in figure~\ref{fig1:b}:
\begin{equation}
\label{eq:oneloopbadSLLScuts}
0 =
\langle \mathcal{L} \, i\, i{+}1 \rangle =
\langle \mathcal{L} \, j{-}1\, j \rangle =
\langle \mathcal{L} \, j\, j{+}1 \rangle =
\langle \mathcal{L} \, k\, k{+}1 \rangle\,.
\end{equation}
These are of three-mass box type and have
the two solutions~\cite{Arkani-Hamed:2013jha}
\begin{equation}
\mathcal{L} = (j\, i\, i{+}1) \cap (j\, k\, k{+}1) \text{ or }
\mathcal{L} = ( \bar{j} \cap (i\, i{+}1), \bar{j} \cap (k\,k{+}1) ).
\end{equation}
The two solutions may be represented respectively by the $D$-matrices
\begin{equation}
D =   \bordermatrix{ & i & i{+}1 & j \cr
& 0 & 0 & 1 \cr
& \langle i{+}1\, j\, k\, k{+}1 \rangle & - \langle i\, j\, k\, k{+}1 \rangle & 0 }
\end{equation}
and
\begin{equation}
D =   \bordermatrix{ & i & i{+}1 & k & k{+}1 \cr
& \langle i{+}1\, \bar{j} \rangle & - \langle i\, \bar{j} \rangle & 0 & 0 \cr
& 0 & 0 & - \langle \bar{j} \, k+1 \rangle &  \langle \bar{j} \, k \rangle}\,.
\end{equation}
Neither matrix is non-negative definite when the $Z$'s are
in the positive domain $G_+(4, n)$, so
we again reach the (correct) conclusion that one-loop MHV
amplitudes do not have singularities on the locus where
eq.~(\ref{eq:oneloopbadSLLS}) is satisfied (for
generic $i$, $j$ and $k$).

\subsection{A Two-Loop Example}

The two-loop scalar double-pentagon integral considered in~\cite{Dennen:2015bet} has a large number of Landau singularities that are spurious singularities of two-loop MHV amplitudes. It would be cumbersome to start with the full list and eliminate the spurious singularities one at a time using the amplituhedron. Here we will be content to consider one example in detail before abandoning this approach in favor of one more directly built on the amplituhedron.

We consider the Landau singularities shown in eq.~(4.12) of~\cite{Dennen:2015bet} which live on the locus
\begin{equation}
\label{eq:twoloopexample}
\langle j \, (j{-}1 \, j{+1}) (i{-}1 \, i) \, (k \, l) \rangle
\langle j\,  (j{-}1 \, j{+}1) (i{-}1 \, i)\,  \bar{k} \cap \bar{l}\rangle = 0\,.	\end{equation}
We consider the generic case when the indices $i$, $j$, $k$, $l$ are
well-separated;
certain degenerate cases do correspond to non-spurious singularities.
This singularity is of pentagon-box type shown in figure~\ref{fig1:c}
since it was found
in~\cite{Dennen:2015bet} to arise from the eight cut conditions
\begin{equation}
\begin{array}{l}
\langle \mathcal{L}^{(1)} \, i{-}1 \, i \rangle   =
\langle  \mathcal{L}^{(1)} \, j{-}1 \, j \rangle   =
\langle \mathcal{L}^{(1)} \, j \, j{+}1 \rangle  =
\langle \mathcal{L}^{(1)} \, \mathcal{L}^{(2} \rangle = 0\,,
\\
\langle  \mathcal{L}^{(2)} \, k{-}1 \, k \rangle   =
\langle   \mathcal{L}^{(2)} \, k \, k{+}1 \rangle   =
\langle  \mathcal{L}^{(2)} \, l{-}1 \, l \rangle   =
\langle  \mathcal{L}^{(2)} \, l \, l{+}1 \rangle = 0\,.
\end{array}
\label{eq:twoloopexamplecuts}
\end{equation}
The last four equations have two solutions $\mathcal{L}^{(2)} = (k \, l)$ or $\mathcal{L}^{(2)} = \bar{k} \cap \bar{l}$, but as in the previous subsection, only the first of these has a chance to avoid being outside the amplituhedron. Taking $\mathcal{L}^{(2)} = (k\, l)$, the two solutions to the first four cut conditions are then
\begin{align}
\mathcal{L}^{(1)} &= (j \, i{-}1 \, i) \cap (j \, k \, l) = (Z_j , Z_{i-1} \langle i \, j \, k \, l \rangle - Z_i \langle i{-}1 \, j \, k \, l \rangle ) \text{ or} \\
\mathcal{L}^{(1)} &= \Big(  (i{-}1 \, i) \cap \bar{j} , (k \, l) \cap \bar{j}      \Big) = \Big( Z_{i-1} \langle i \, \bar{j}  \rangle - Z_{i} \langle i{-}1 \, \bar{j} \rangle , Z_k \langle l \bar{j} \rangle - Z_l \langle k \, \bar{j} \rangle  \Big)\,.
\end{align}
The $D$-matrices corresponding to the first solution can be taken as
\begin{equation}
\begin{pmatrix}
D^{(1)} \cr
D^{(2)}
\end{pmatrix} =   \bordermatrix{ & i{-}1 & i & j & k & l \cr
& 0 & 0  & 1 & 0 & 0 \cr
& \langle i \, j \, k \, l \rangle &-  \langle i{-}1 \, j \, k \, l  \rangle & 0 & 0 & 0 \cr
& 0 & 0 & 0 & 1 & 0 \cr
& 0 & 0 & 0 & 0 & 1 }\,.
\end{equation}
Evidently two of its $4 \times 4$ minors are $- \langle i \, j \, k \, l \rangle$ and $\langle i{-}1 \, j \, k \, l \rangle$, which have opposite signs for generic $Z$ in the positive domain. $D$-matrices corresponding to the second solution can be written as
\begin{equation}
\begin{pmatrix}
D^{(1)} \cr
D^{(2)}
\end{pmatrix}
=   \bordermatrix{ & i{-}1 & i  & k & l \cr
& \langle i \, \bar{j} \rangle & - \langle i{-}1 \, \bar{j} \rangle  & 0 & 0 \cr
& 0 & 0  & \langle l \bar{j} \rangle & - \langle k \, \bar{j} \rangle \cr
& 0 & 0 &  1 & 0 \cr
& 0 & 0 &  0 & 1 }\,,
\end{equation}
which again has minors of opposite signs.

We conclude that the locus where the cut conditions~(\ref{eq:twoloopexamplecuts}) are satisfied lies strictly outside the amplituhedron, and therefore that there is no discontinuity around the putative branch point at~(\ref{eq:twoloopexample}). Indeed, this is manifested by the known fact~\cite{CaronHuot:2011ky} that two-loop MHV amplitudes do not have symbol entries which vanish on this locus. Actually, while correct, we were slightly too hasty in reaching this conclusion, since we only analyzed one set of cut conditions.  Although it doesn't happen in this example, in general there may exist several different collections of cut conditions associated to the same Landau singularity, and the discontinuity around that singularity would receive additive contributions from each distinct set of associated cut contributions.

\subsection{Summary}

We have shown, via a slight refinement of the analysis carried out
in~\cite{Dennen:2015bet}, that the spurious branch points of one- and
two-loop MHV amplitudes encountered in that paper can be eliminated
simply on the basis of positivity constraints in the amplituhedron.
It is simple to see that the cuts considered above have no support
for MHV amplitudes so it may seem like overkill to use the
fancy language of the amplituhedron.
However we wanted to highlight the following approach:

(1) First, consider a representation of an amplitude as a sum over a particular type of Feynman integrals. Find the Landau singularities of a generic term in the sum. These tell us the loci in $Z$-space where the amplitude {\it may}
have a singularity.

(2) For each {\it potential} singularity obtained in (1), check whether the corresponding on-shell conditions have a non-zero intersection with the (closure of) the amplituhedron. If the answer is no, for all possible sets of cut conditions associated with a given Landau singularity, then the singularity must be spurious.

This approach is conceptually straightforward but inefficient.
One manifestation of this inefficiency is that although double
pentagon integrals are characterized by four free indices $i$, $j$,
$k$, $l$, we will see in the next section the vast majority
of the resulting potential singularities are spurious.
Specifically we will see that in order for the solution to a given
set of cut conditions to have support inside the (closure of the)
amplituhedron, the conditions must be relaxed in such a way that
they involve only three free indices.
In other words, most of the $\mathcal{O}(n^4)$ singularities
of individual double pentagon integrals must necessarily cancel
out when they are summed,
leaving only $\mathcal{O}(n^3)$ physical singularities
of the full two-loop MHV amplitudes.
(The fact that these amplitudes have only $\mathcal{O}(n^3)$
singularities is manifest in the result of~\cite{CaronHuot:2011ky}.)
This motivates us to seek a more ``amplituhedrony" approach to finding singularities where we do not start by considering any particular representation of the amplitude, but instead start by thinking directly about positive configurations of loops $\mathcal{L}^{(\ell)}$.

\section{An Amplituhedrony Approach}
\label{sec:amplituhedrony}

The most significant drawback of the approach taken in the previous section is that it relies on having explicit representations of an integrand in terms of local Feynman integrals.
These have been constructed for all two-loop amplitudes
in SYM theory~\cite{Bourjaily:2015jna}, but at higher loop
order even finding such representations becomes a huge computational
challenge that we would like to be able to bypass.
Also, as the loop order increases, the number of potential Landau singularities grows rapidly, and the vast majority of these potential singularities will fail the positivity analysis and hence turn out to be spurious.
We would rather not have to sift through all of this chaff to find the wheat.

Let's begin by taking a step back to appreciate that the only reason we
needed the crutch of local Feynman integrals
in the previous section
is that each Feynman diagram topology provides
a set of propagators
for which we can solve the associated Landau
equations~(\ref{eq:LS1}) and~(\ref{eq:LS2}) to find
potential singularities.
Then, for each set of cut conditions, we
can determine whether the associated Landau singularity is physical
or spurious by asking the amplituhedron whether or not the set of loops
$\mathcal{L}^{(\ell)}$ satisfying the cut conditions has any overlap
with the amplituhedron.

In this section we propose a more ``amplituhedrony'' approach that does not rely on detailed knowledge of integrands.
We invert the logic of the previous section:
instead of using
Feynman diagrams to generate sets of cut conditions that we need
to check one by one,
we can ask the amplituhedron itself to directly identify all
potentially ``valid'' sets of cut conditions that
are possibly relevant to the singularities of
an amplitude.

To phrase the problem more abstractly:
for a planar $n$-particle amplitude at $L$-loop order, there are in general
$n L + L(L-1)/2$ possible local cut conditions one can write down:
\begin{equation}
\label{eq:gendom}
\langle \mathcal{L}^{(\ell)}\, i\, i{+}1 \rangle = 0 \text{ for all } \ell, i
\text{ and } \langle \mathcal{L}^{(\ell_1)} \, \mathcal{L}^{(\ell_2)} \rangle
= 0 \text{ for all } \ell_1 \ne \ell_2.
\end{equation}
We simply need to characterize which subsets of these cut
conditions can possibly be simultaneously satisfied for loop momenta
$\mathcal{L}^{(\ell)}$ living in the closure of the
amplituhedron.
Each such set of cut conditions is a subset of one or more
maximal subsets, and these maximal subsets are just the maximal
codimension boundaries of the amplituhedron.

Fortunately,
the maximal codimension boundaries of the MHV amplituhedron are particularly
simple, as explained in~\cite{Arkani-Hamed:2013kca}.
Each loop momentum $\mathcal{L}^{(\ell)}$ must take the form
$(i \, j)$ for some $i$ and $j$ (that can be different for different
$\ell$),
and the condition of mutual positivity enforces an
emergent planarity:  if all of the lines $\mathcal{L}^{(\ell)}$
are drawn as chords
on a disk between points on the boundary
labeled $1, 2, \ldots, n$,
then positivity forbids any two lines to cross in the interior of the disk.
In what follows we follow a somewhat low-brow analysis in which we
systematically consider relaxations away from the maximum
codimension boundaries, but the procedure can be streamlined
by better harnessing this emergent planarity, which certainly
pays off at higher loop order~\cite{TD}.

In the next few subsections we demonstrate
this ``amplituhedrony'' approach
explicitly at one and two loops before summarizing the main
idea at the end of the section.

\subsection{One-Loop MHV Amplitudes}

\begin{figure}[]
\centering
\hspace*{\fill}
\subcaptionbox{\label{fig2:a}}{\includegraphics[scale=1.44]{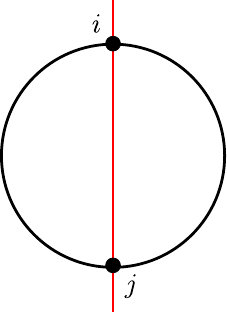}}
\hfill
\subcaptionbox{\label{fig2:b}}{\raisebox{0.9cm}{\includegraphics[scale=1.44]{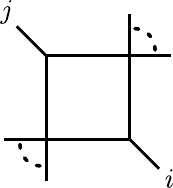}}}
\hspace*{\fill}
\caption{(a) A maximum codimension boundary of the one-loop
MHV amplituhedron.  The circle is a schematic depiction of
the $n$ line segments $(1 \, 2)$, $(2 \, 3)$, \ldots, $(n\, 1)$
connecting the $n$ cyclically ordered external kinematic
points $Z_i \in G_+(4,n)$ and the red line shows
the loop momentum $\mathcal{L} = (i\, j)$.
(b) The corresponding Landau diagram, which is a graphical depiction
of the four cut conditions~(\ref{eq:oneloopcuts}) that are
satisfied on this boundary.}
\label{fig2}
\end{figure}

The maximum codimension boundaries of the one-loop MHV
amplituhedron occur when
\begin{equation}
\mathcal{L} = (i\, j)\,,
\end{equation}
as depicted in figure~\ref{fig2:a}.
On this boundary four cut conditions of
``two-mass easy'' type~\cite{Bern:1994zx} are manifestly satisfied:
\begin{equation}
\langle \mathcal{L} \, i{-}1 \, i \rangle =
\langle \mathcal{L} \, i \, i{+}1 \rangle =
\langle \mathcal{L} \, j{-}1 \, j \rangle =
\langle \mathcal{L} \, j \, j{+}1 \rangle = 0\,,
\label{eq:oneloopcuts}
\end{equation}
as depicted in the Landau diagram shown in figure~\ref{fig2:b}.
(For the moment we consider $i$ and $j$ to be well separated so there
are no accidental degenerations.)
The Landau analysis of eq.~(\ref{eq:oneloopcuts})
has been performed long ago~\cite{Landau:1959fi,ELOP} and
reviewed in the language of momentum twistors
in~\cite{Dennen:2015bet}. A leading solution to the Landau
equations exists only if
\begin{equation}
\langle i \, \bar{j} \rangle \langle \bar{i} \, j \rangle = 0\,.
\end{equation}

Subleading
Landau equations are obtained
by relaxing one of the four on-shell conditions.
This leads to cuts of two-mass triangle type, which are uninteresting
(they exist for generic kinematics, so don't correspond to
branch points of the amplitude).
At sub-subleading order we reach cuts of bubble type. For example
if we relax the second and fourth condition in eq.~(\ref{eq:oneloopcuts})
then we encounter a Landau singularity which lives on the locus
\begin{equation}
\langle i{-}1\, i\, j{-}1\, j \rangle = 0\,.
\label{eq:bubble}
\end{equation}
Other relaxations either give no constraint on kinematics, or the same
as eq.~(\ref{eq:bubble}) with $i \to i{+}1$ and/or $j \to j{+}1$.

Altogether, we reach the conclusion that all physical branch points
of one-loop MHV amplitudes occur on loci of the form
\begin{equation}
\label{eq:oneloopalphabet}
\langle a \, \bar{b} \rangle = 0 \text{ or } \langle
a \, a{+}1 \, b \, b{+}1 \rangle = 0
\end{equation}
for various $a, b$.
(Note that whenever we say there is a branch point at $x=0$,
we mean more specifically that there is a branch cut between $x=0$
and $x=\infty$.)
Indeed, these exhaust the branch points of the one-loop
MHV amplitudes (first computed in~\cite{Bern:1994zx})
except for branch points arising as a consequence of infrared
regularization, which are captured by the BDS
ansatz~\cite{Bern:2005iz}.

\subsection{Two-Loop MHV Amplitudes: Configurations of Positive Lines}

We divide the two-loop analysis into two steps.  First, in this subsection,
we classify valid configurations of mutually non-negative lines.
This provides a list of the sets of cut conditions on which two-loop
MHV amplitudes have nonvanishing support.
Then in the following subsection we solve the Landau equations for each
set of cut conditions, to find the actual location of the corresponding
branch point.

At two loops the MHV amplituhedron
has two distinct kinds of maximum
codimension boundaries~\cite{Arkani-Hamed:2013kca}.
The first type has
$\mathcal{L}^{(1)} = (i\, j)$ and $\mathcal{L}^{(2)} = (k\, l)$
for distinct cyclically ordered $i$, $j$, $k$, $l$.
Since $\langle \mathcal{L}^{(1)} \, \mathcal{L}^{(2)} \rangle$
is non-vanishing (inside the positive domain $G_+(4,n)$) in this case,
this boundary can be thought of as corresponding to a cut of a
product of one-loop Feynman integrals, with no common propagator
$\langle \mathcal{L}^{(1)} \, \mathcal{L}^{(2)} \rangle$.
Therefore we will not learn
anything about two-loop singularities beyond what is already
apparent at one loop.

\begin{figure}[]
\centering
\hspace*{\fill}
\subcaptionbox{\label{fig3:a}}{\raisebox{0.3cm}{\includegraphics[scale=1.44]{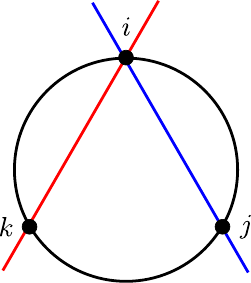}}}
\hfill
\subcaptionbox{\label{fig3:b}}{\includegraphics[scale=1.44]{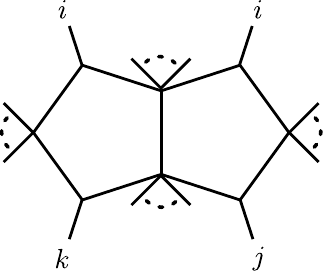}}
\hspace*{\fill}
\caption{(a) A maximum codimension boundary of the two-loop
MHV amplituhedron.
(b) The corresponding Landau diagram (which, it should be
noted, does not have the form of a standard Feynman integral)
depicting the nine
cut conditions~(\ref{eq:firstfour})--(\ref{eq:last}) that are
satisfied on this boundary.}
\label{fig3}
\end{figure}

The more interesting type of maximum codimension boundary
has $\mathcal{L}^{(1)} = (i \, j)$ and $\mathcal{L}^{(2)} = (i \, k)$,
as depicted in figure~\ref{fig3:a}.
Without loss of generality $i < j < k$, and for now we will moreover
assume that $i$, $j$ and $k$ are well-separated to avoid
any potential degenerations.
(These can be relaxed at the end of the analysis, in particular
to see that the degenerate case $j=k$ gives nothing interesting.)
On this boundary the following nine cut conditions shown in the
Landau diagram of figure~\ref{fig3:b} are
simultaneously satisfied:
\begin{align}
\label{eq:firstfour}
\langle \mathcal{L}^{(1)} \, i{-}1 \, i \rangle   =
\langle \mathcal{L}^{(1)} \, i \, i{+}1 \rangle   =
\langle \mathcal{L}^{(2)} \, i{-}1 \, i \rangle  =
\langle \mathcal{L}^{(2)} \, i \, i{+}1 \rangle = 0\,,
\\
\label{eq:nextfour}
\langle \mathcal{L}^{(1)} \, j{-}1 \, j \rangle   =
\langle \mathcal{L}^{(1)} \, j \, j{+}1 \rangle   =
\langle \mathcal{L}^{(2)} \, k{-}1 \, k \rangle   =
\langle \mathcal{L}^{(2)}\, k \, k{+}1 \rangle = 0\,,
\\
\langle \mathcal{L}^{(1)} \, \mathcal{L}^{(2)} \rangle = 0\,.
\label{eq:last}
\end{align}
This is the maximal set of cuts that can be simultaneously
satisfied while keeping the $\mathcal{L}^{(\ell)}$'s inside the
closure of the amplituhedron for generic $Z \in G_+(4, n)$.
We immediately note that since only three free indices $i$, $j$, $k$
are involved, this set of cuts manifestly has size
$\mathcal{O}(n^3)$, representing
immediate savings compared to the larger $\mathcal{O}(n^4)$
set of double-pentagon cut conditions as discussed at the end of the
previous section.

We can generate other, smaller
sets of cut conditions by relaxing some of the nine shown in
eqs.~(\ref{eq:firstfour})--(\ref{eq:last}).
This corresponds to looking at subleading singularities, in the language
of the Landau equations.
However,
it is not interesting to consider relaxations that
lead to
$\langle \mathcal{L}^{(1)} \, \mathcal{L}^{(2)} \rangle \ne 0$
because, as mentioned above, it essentially
factorizes the problem into a product of one-loop cuts.
Therefore
in what follows we only consider cuts
on which $\langle \mathcal{L}^{(1)} \, \mathcal{L}^{(2)} \rangle = 0$.

By relaxing various subsets of the other 8 conditions we can generate $2^8$ subsets of cut conditions.  In principle each subset should be analyzed
separately, but there is clearly a natural stratification of
relaxations which we can exploit to approach the problem
systematically.
In fact, we will see that the four cut
conditions in eq.~(\ref{eq:firstfour}) that involve the point $i$
play a special role.
Specifically, we will see that the four cut conditions in
eq.~(\ref{eq:nextfour}) involving $j$ and $k$
can always be relaxed, or un-relaxed, ``for
free'', with no impact on positivity.
Therefore, we see that whether a configuration of loops may be positive
or not depends only on which subset of the
four cut conditions~(\ref{eq:firstfour})
is relaxed.

In this subsection we will classify the subsets of eq.~(\ref{eq:firstfour})
that lead to valid configurations of positive lines $\mathcal{L}^{(\ell)}$,
and in the next subsection we will find the locations of the corresponding
Landau singularities.

\paragraph{Relaxing none of eq.~(\ref{eq:firstfour}) [figure~\ref{fig3:a}].}

At maximum codimension
we begin with the obviously valid pair
of mutually non-negative lines represented trivially by
\begin{equation}
\begin{pmatrix}
D^{(1)} \cr
D^{(2)}
\end{pmatrix}
= \bordermatrix{ & i & j & k \cr
 & 1 & 0 & 0 \cr
 & 0  & 1 & 0 \cr
 & 1 & 0 & 0 \cr
 & 0 & 0 & 1 }\,.
\end{equation}

\begin{figure}[]
\centering
\subcaptionbox{\\
$\langle\color{red}\mathcal{L}^{(2)}\color{black} \, i\, i{+}1\rangle
\ne 0$\label{fig4:a}}[5.2cm]{\includegraphics[scale=1.44]{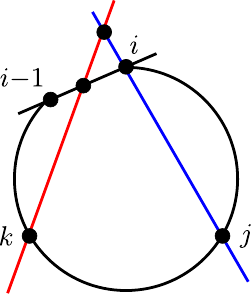}}
\hspace{-0.35cm}
\subcaptionbox{\\
$\langle\color{red}\mathcal{L}^{(2)}\color{black} \, i\, i{+}1\rangle,
\langle \color{blue}\mathcal{L}^{(1)}\color{black} \, i{-}1\, i \rangle
\ne 0$\label{fig4:b}}[5.3cm]{\includegraphics[scale=1.44]{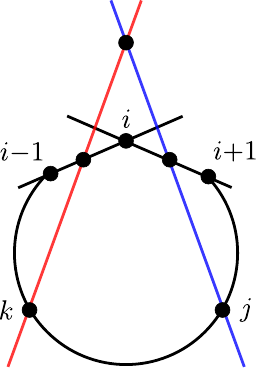}}
\hspace{-0.35cm}
\subcaptionbox{\\
$\langle\color{blue}\mathcal{L}^{(1)}\color{black} \, i\, i{+}1\rangle,
\langle \color{red}\mathcal{L}^{(2)}\color{black} \, i{-}1\, i \rangle
\ne 0$\label{fig4:c}}[5.2cm]{\includegraphics[scale=1.44]{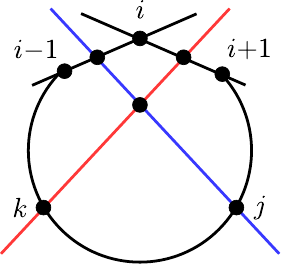}}
\caption{Three different invalid relaxations
of the maximal codimension boundary shown in figure~\ref{fig3}.}
\label{fig4}
\end{figure}

\paragraph{Relaxing any one of eq.~(\ref{eq:firstfour}).}

The four cases are identical up to relabeling so we
consider
relaxing the condition
$\langle \mathcal{L}^{(2)}\, i\, i{+}1 \rangle = 0$,
shown in figure~\ref{fig4:a}. In this case
the remaining seven cut conditions
on the first two lines of eqs.~(\ref{eq:firstfour}) and~(\ref{eq:nextfour})
admit the one-parameter
family of solutions
\begin{equation}
\mathcal{L}^{(1)} = (i\, j), \qquad
\mathcal{L}^{(2)} = ( Z_k, \alpha Z_{i-1} + (1 - \alpha) Z_i)\,.
\end{equation}
We recall that the parity conjugate solutions
having $\mathcal{L}^{(1)} = \bar{i} \cap \bar{j}$ lie discretely
outside the amplituhedron as seen in eq.~(\ref{eq:bad}).
The corresponding $D$-matrices
\begin{equation}
\begin{pmatrix}
D^{(1)} \cr
D^{(2)}
\end{pmatrix}
= \bordermatrix{ & i-1 & i & j & k \cr
& 0 & 1 & 0 & 0 \cr
& 0 & 0  & 1 & 0 \cr
& \alpha & 1-\alpha & 0 & 0 \cr
& 0 & 0 & 0 & 1 }
\end{equation}
are mutually non-negative for $0 \le \alpha \le 1$.
It remains to impose the final cut condition that $\mathcal{L}^{(1)}$ and
$\mathcal{L}^{(2)}$ intersect:
\begin{equation}
\langle \mathcal{L}^{(1)}  \mathcal{L}^{(2)} \rangle  =
 \alpha \langle i{-}1 \, i\, j \,k \rangle = 0\,.
\end{equation}
For general positive external kinematics
this will only be satisfied when $\alpha = 0$, which brings us back to the maximum codimension boundary.
We conclude that the loop configurations of this type do not
generate branch points.

\paragraph{Relaxing
$\langle \mathcal{L}^{(1)} \, i{-}1\,i \rangle = 0$
and
$\langle \mathcal{L}^{(2)} \, i\, i{+}1 \rangle = 0$ [figure~\ref{fig4:b}].}

In this case the six remaining cut conditions in eqs.~(\ref{eq:firstfour})
and~(\ref{eq:nextfour})
admit the two-parameter family of solutions
\begin{equation}
\mathcal{L}^{(1)} = (\alpha Z_i + (1-\alpha) Z_{i+1} , Z_j), \quad
\mathcal{L}^{(2)} = (\beta Z_i + (1-\beta) Z_{i-1} , Z_k)\,.
\end{equation}
The corresponding $D$-matrices
\begin{equation}
\begin{pmatrix}
D^{(1)} \cr
D^{(2)}
\end{pmatrix}
= \bordermatrix{& i-1 & i & i+1 & j & k \cr
&  0 & \alpha & 1-\alpha & 0 & 0 \cr
& 0 & 0 & 0 & 1 & 0 \cr
& 1-\beta& \beta  & 0 & 0 & 0 \cr
& 0 & 0 & 0 & 0 & 1 }
\end{equation}
are mutually non-negative if $0 \le \alpha, \beta \le 1$.
Imposing that the two loops intersect gives the constraint
\begin{equation}
\langle \mathcal{L}^{(1)} \mathcal{L}^{(2)}\rangle
= \alpha (1-\beta) \langle i{-}1 \, i \, j \, k \rangle + (1-\alpha) \beta \langle i \, i{+}1 \, j \, k \rangle + (1-\alpha) (1-\beta) \langle i{-}1 \, i{+}1 \, j \, k \rangle = 0\,,
\end{equation}
which is not satisfied for general positive kinematics unless
$\alpha = \beta = 1$, which again brings us back to the maximum
codimension boundary.

Relaxing the two conditions
$\langle \mathcal{L}^{(1)} \, i \, i{+}1 \rangle = \langle \mathcal{L}^{(2)} \, i \, i{-}1 \rangle = 0$,
depicted in figure~\ref{fig4:c}, is easily
seen to lead to the same conclusion.

\begin{figure}[]
\centering
\hspace*{\fill}
\subcaptionbox{\\
$\langle\color{red}\mathcal{L}^{(2)}\color{black} \, i\, i{+}1\rangle,
\langle \color{blue}\mathcal{L}^{(1)}\color{black} \, i\, i{+}1 \rangle
\ne 0$\label{fig5:a}}[5.3cm]{\includegraphics[scale=1.44]{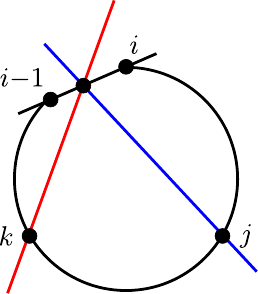}}
\hfill
\subcaptionbox{\\
$\langle\color{red}\mathcal{L}^{(2)}\color{black} \, i{-}1\, i\rangle,
\langle \color{red}\mathcal{L}^{(2)}\color{black} \, i\, i{+}1 \rangle
\ne 0$\label{fig5:b}}[5.3cm]{\raisebox{0.02cm}{\includegraphics[scale=1.44]{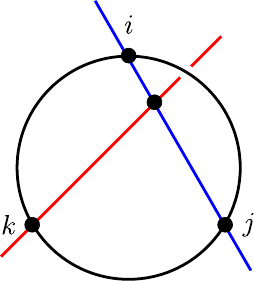}}}
\hspace*{\fill}
\caption{Two valid double relaxations of figure~\ref{fig3}. The
other two possibilities are obtained by taking $i \to i{+}1$ in~(a)
or $\color{red}\mathcal{L}^{(2)}\color{black} \to \color{blue}
\mathcal{L}^{(1)}\color{black}$ and $j \leftrightarrow k$ in~(b).}
\label{fig5}
\end{figure}

\paragraph{Relaxing
$\langle \mathcal{L}^{(1)} \, i \, i{+}1 \rangle=0$ and $\langle \mathcal{L}^{(2)} \, i \, i{+}1 \rangle=0$~[figure~\ref{fig5:a}].}

In this case
there is a one-parameter family of solutions
satisfying all seven
remaining cut conditions including
$\langle \mathcal{L}^{(1)}\, \mathcal{L}^{(2)} \rangle =0$:
\begin{equation}
\mathcal{L}^{(1)} = ( \alpha Z_i + (1 - \alpha) Z_{i + 1},
Z_j), \qquad
\mathcal{L}^{(2)} = ( \alpha Z_i + (1 - \alpha) Z_{i + 1}, Z_k)\,.
\end{equation}
The $D$-matrices can be represented as
\begin{equation}
\begin{pmatrix}
D^{(1)} \cr
D^{(2)}
\end{pmatrix}
=   \bordermatrix{ & i & i{+}1 & j & k \cr
&  \alpha & 1-\alpha & 0 & 0  \cr
& 0 & 0 & 1 & 0 \cr
& \alpha & 1 - \alpha & 0 & 0 \cr
& 0 & 0 & 0 & 1 }\,,
\end{equation}
which is a valid mutually non-negative configuration for
$0 \le \alpha \le 1$.
We conclude that these configurations represent physical
branch points of two-loop MHV amplitudes by appealing to Cutkoskian
intuition, according to which we would compute the discontinuity of
the amplitude around this branch point
by integrating over $0 \le \alpha \le 1$ (in
figure~\ref{fig5:a} this corresponds to integrating
the intersection point
of the two $\mathcal{L}$'s
over the line segment between $Z_{i-1}$ and $Z_i$).

Relaxing the two conditions
$\langle \mathcal{L}^{(1)} \, i \, i{-}1 \rangle = \langle \mathcal{L}^{(2)} \, i \, i{-}1 \rangle = 0$ is clearly equivalent up to relabeling.

\paragraph{Relaxing
$\langle\mathcal{L}^{(2)}\, i{-}1 \, i \rangle = 0$ and $\langle\mathcal{L}^{(2)} \, i\, i{+}1 \rangle = 0$ [figure~\ref{fig5:b}].}

The seven remaining cut conditions admit a one-parameter family of solutions
\begin{equation}
\mathcal{L}^{(1)} = (i\,j), \qquad
\mathcal{L}^{(2)} = (\alpha Z_i + (1 - \alpha) Z_j, Z_k)\,,
\end{equation}
which can be represented by
\begin{equation}
\begin{pmatrix}
D^{(1)} \cr
D^{(2)}
\end{pmatrix}
=   \bordermatrix{ & i & j & k \cr
&  1 & 0 & 0  \cr
& 0 & 1 & 0   \cr
& \alpha & 1 - \alpha & 0  \cr
& 0 & 0 & 1 }\,.
\end{equation}
This is a valid configuration of mutually non-negative lines
for $0 \le \alpha
\le 1$ so we expect it to correspond to a physical
branch point.
Clearly the same conclusion holds if
we were to completely relax $\mathcal{L}^{(1)}$ at $i$ instead of
$\mathcal{L}^{(2)}$.

\paragraph{Higher relaxations of eq.~(\ref{eq:firstfour}).}

So far we have considered the relaxation of any one or any two of
the conditions shown in eq.~(\ref{eq:firstfour}).
We have found that single relaxations do not yield
branch points of the amplitude, and that four of the
six double relaxations are valid while the two
double relaxations
shown in figures~\ref{fig4:b}
and~\ref{fig4:c} are invalid.

What about triple relaxations?
These can be checked by explicit construction of the
relevant $D$-matrices, but it is also easy to see graphically that
any triple relaxation is valid because they
can all be reached by relaxing one of the valid
double relaxations. For example, the triple relaxation where
we relax all of eq.~(\ref{eq:firstfour}) except
$\langle \mathcal{L}^{(1)}\, i{-}1\, i \rangle = 0$ can be realized
by rotating $\mathcal{L}^{(2)}$ in figure~\ref{fig5:a}
clockwise around the point $k$ so that it continues to
intersect $\mathcal{L}^{(1)}$.
As a second example, the triple relaxation
where we relax all but $\langle \mathcal{L}^{(2)} \, i{-}1\, i\rangle = 0$
can be realized by rotating $\mathcal{L}^{(1)}$ in figure~\ref{fig5:a}
counter-clockwise around the point $j$ so that it continues
to intersect $\mathcal{L}^{(2)}$.

Finally we turn to the case when all four cut
conditions in eq.~(\ref{eq:firstfour}) are relaxed.
These relaxed cut conditions admit two branches of solutions,
represented by $D$-matrices of the form
\begin{equation}
\begin{pmatrix}
D^{(1)} \cr
D^{(2)}
\end{pmatrix}
=   \bordermatrix{ & j & j+1 & \cdots & k-1 & k \cr
&  1 & 0 & \cdots & 0 & 0 \cr
& \alpha_j & \alpha_{j+1} & \cdots & \alpha_{k-1} & \alpha_k   \cr
& \alpha_j & \alpha_{j+1}  & \cdots & \alpha_{k-1} & \alpha_k \cr
& 0 & 0 & \cdots & 0 & 1 }
\end{equation}
or a similar form with $\alpha$ parameters wrapping the other way
around from $k$ to $j$:
\begin{equation}
\begin{pmatrix}
D^{(1)} \cr
D^{(2)}
\end{pmatrix}
= \bordermatrix{& \cdots & j{-}1 & j & k & k{+}1 & \cdots\cr
&\cdots &  \alpha_{j-1} &  \alpha_j & - \alpha_k & - \alpha_{k+1}
& \cdots  \cr
&\cdots & 0  &  1 & 0 & 0 & \cdots \cr
&\cdots &  \alpha_{j-1} &  \alpha_j & - \alpha_k & - \alpha_{k+1}
& \cdots \cr
&\cdots & 0 & 0 &1& 0 & \cdots } \,.
\label{eq:under}
\end{equation}
Both of these parameterize valid configuration of mutually non-negative lines
as long as all of the $\alpha$'s are positive.

\paragraph{Relaxing $\mathcal{L}^{(1)}$ at $j$ and/or $\mathcal{L}^{(2)}$ at
$k$.}

All of the configurations we have considered so far
keep the four propagators in eq.~(\ref{eq:nextfour}) on shell.
However it is easy to see that none of these conditions
have any bearing on positivity one way or the other.
For example, there is no way to
render the configuration shown in
figure~\ref{fig4:b} positive by moving
$\mathcal{L}^{(1)}$ away from the vertex $j$
while maintaining all of the other cut conditions.
On the other hand, there is no way to spoil the positivity
of the configuration shown in figure~\ref{fig5:b}
by moving $\mathcal{L}^{(2)}$ away from the vertex $k$ while
maintaining all other cut conditions.

\paragraph{Summary.}

We call a set of cut conditions ``valid'' if the
$m \ge 0$-dimensional
locus in $\mathcal{L}$-space where the conditions are simultaneously
satisfied has non-trivial $m$-dimensional overlap with the closure
of the amplituhedron.  (The examples shown in figures~\ref{fig5:a}
and~\ref{fig5:b} both have $m=1$, but further
relaxations would have higher-dimensional solution spaces.)
As mentioned above, this criterion is motivated by Cutkoskian
intuition that the discontinuity of the amplitude would be computed
by an integral over the intersection of this locus with the (closure of the)
amplituhedron.  If this intersection is empty (or lives
on a subspace that is less than $m$-dimensional) then such an integral
would vanish, signalling that the putative singularity is actually
spurious.

The nine cut conditions shown in eqs.~(\ref{eq:firstfour})--(\ref{eq:last})
are solved
by the configuration of lines shown in figure~\ref{fig3:a} that is
a zero-dimensional boundary of the amplituhedron.
We have systematically investigated relaxing various
subsets of these conditions (with the exception
of eq.~(\ref{eq:last}), to stay
within the realm of genuine two-loop singularities)
to determine which relaxations are ``valid'' in the sense just described.

\paragraph{Conclusion:}
The most general valid relaxation of the configuration
shown in figure~\ref{fig3:a} is either an arbitrary relaxation
at the points $j$ and $k$, or an arbitrary relaxation
of figure~\ref{fig5:a} (or the same with $i \mapsto i{+}1$),
or an arbitrary relaxation of figure~\ref{fig5:b} (or
the same with $j \leftrightarrow k$).
The configurations shown in figure~\ref{fig4}, and further
relaxations thereof that are not relaxations of
those shown in figure~\ref{fig5}, are invalid.

\subsection{Two-Loop MHV Amplitudes: Landau Singularities}
\label{sec:33}

\begin{figure}[]
\centering
\hspace*{\fill}
\subcaptionbox{\label{fig6:a}}{\raisebox{0.25cm}{\includegraphics[scale=1.44]{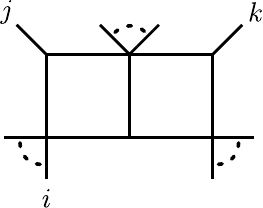}}}
\hfill
\subcaptionbox{\label{fig6:b}}{\includegraphics[scale=1.44]{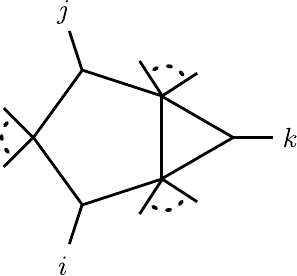}}
\hspace*{\fill}
\caption{The Landau diagrams showing the seven cut conditions
satisfied by figures~\ref{fig5:a} and~\ref{fig5:b}, respectively.}
\end{figure}

In the previous subsection we asked the amplituhedron directly
to tell us which possible sets of cut conditions are valid for
two-loop MHV amplitudes, rather than starting from some integral
representation and using the amplituhedron to laboriously sift
through the many
spurious singularities.
We can draw Landau diagrams for each valid relaxation to serve as a
graphical indicator of the cut conditions that are satisfied.
The Landau diagram with nine propagators corresponding to the
nine cut conditions satisfied by figure~\ref{fig3:a}
was already displayed in
figure~\ref{fig3:b}.
The configurations shown in figures~\ref{fig5:a}
and~\ref{fig5:b} satisfy the seven cut conditions
corresponding to the seven propagators in
figures~\ref{fig6:a}
and~\ref{fig6:b}, respectively.
We are now ready to determine the locations of the branch points
associated to these valid cut configurations (and their
relaxations) by solving the Landau
equations.

The following calculations follow very closely
those done in~\cite{Dennen:2015bet}.
Note that throughout this section, in solving cut conditions we will
always ignore branches of solutions (for example
those of the type $\mathcal{L} = \bar{i} \cap \bar{j}$) which
cannot satisfy positivity.

\paragraph{The double-box.}

For the double-box shown in in figure~\ref{fig6:a}
let us use $A \in \mathbb{P}^3$ to denote the point
on the line $(i{-}1,i)$
where the two loop lines $\mathcal{L}^{(\ell)}$
intersect.  These can then be parameterized
as $\mathcal{L}^{(1)} = (A, Z_j)$ and $\mathcal{L}^{(2)} = (A, Z_k)$.
The quickest way to find the location of the
leading Landau singularity is to
impose eq.~(\ref{eq:LS1}) for each of the two loops.
These are both of two-mass easy type, so we find
that the Landau singularity lives on the locus
(see~\cite{Dennen:2015bet})
\begin{equation}
\langle i{-}1\, i\, j \, k \rangle \langle A \, \bar{j} \rangle =
\langle i{-}1\, i \, j \, k \rangle \langle A \, \bar{k} \rangle = 0\,.
\end{equation}
These can be solved in two ways; either
by
\begin{equation}
\label{eq:ok1}
\langle i{-}1\, i\, j\, k \rangle = 0
\end{equation}
or by solving the first condition for $A = \bar{j} \cap (i{-}1\, i)$ and
substituting this into the second condition to find
\begin{equation}
\langle i{-}1\,i\, \bar{j} \cap \bar{k} \rangle = 0\,.
\end{equation}

The astute reader may recall that in~(\ref{eq:oneloopbadLLSlocus})
we discarded a singularity
of the same type as in eq.~(\ref{eq:ok1}).
This example highlights that it is crucial to appreciate
the essential asymmetry between the roles of the two
types of Landau equations.
The on-shell conditions~(\ref{eq:LS1}) by themselves only provide
information about {\it{discontinuities}}.   We discarded
eq.~(\ref{eq:oneloopbadLLSlocus}) because
the solution
has support on a set of measure zero inside the closure of the
amplituhedron, signalling that there
is no discontinuity around the branch cut associated to the cut
conditions shown in eq.~(\ref{eq:oneloopbadLLS}).
Therefore we never needed to inquire as to the actual location where the
corresponding branch point
might have been.
To learn about the {\it{location}} of a branch point we have to solve also
the second type of Landau
equations~(\ref{eq:LS2}).
Indeed~(\ref{eq:ok1}) does correspond to a branch point that lies outside
the positive domain, but we don't discard it because
the discontinuity of the amplitude around this branch point
is nonzero.  As mentioned above, according to the Cutkosky rules it would
be computed by an integral over the line segment
between $Z_{i-1}$ and $Z_i$ in figure~\ref{fig5:a}.
When branch points lie outside $G_+(4,n)$, as in this case, it signals a
discontinuity that does not exist on the physical sheet
but on some other sheet; see the comments near the end of
section~1.

Additional (sub${}^k$-leading, for various $k$) Landau singularities
are exposed by
setting various sets of $\alpha$'s to zero in the Landau equations
and relaxing
the associated cut conditions.
Although these precise configurations
were not analyzed in~\cite{Dennen:2015bet},
the results of that paper, together with some very useful
tricks reviewed in appendix~A, are easily used to reveal
branch points at the loci
\begin{equation}
\langle j(j{-}1,j{+}1)(k,k{\pm}1)(i{-}1,i)\rangle = 0\,
\end{equation}
together with
the same for $j \leftrightarrow k$, as well as
$\langle a\, a{+}1\, b\, b{+}1 \rangle = 0$
for $a$, $b$ drawn
from the set $\{i{-}1, j{-}1, j, k{-}1, k\}$.

\paragraph{The pentagon-triangle.}

With the help of appendix~A and the results of~\cite{Dennen:2015bet}
it is easily seen that the
leading singularity of the
pentagon-triangle shown in figure~\ref{fig6:b}
is located on the locus where
\begin{equation}
\langle i \bar{j} \rangle \langle \bar{i} j \rangle = 0\,.
\end{equation}
The computation of additional singularities essentially
reduces to the same calculation for a three-mass pentagon,
which was carried out in~\cite{Dennen:2015bet}. Altogether
we find
that branch points live on the loci
\begin{align}
\begin{split}
\langle i \, j \, k{-} 1\, k \rangle = 0\,, \\
\langle i (i{-}1\, i{+}1) (j{-}1\, j) (k{-}1\, k) \rangle = 0\,, \\
\langle i (i{-}1\, i{+}1) (j\, j{+}1) (k{-}1\, k) \rangle = 0\,, \\
\langle j (j{-}1\, j{+}1) (i{-}1\, i) (k{-}1\, k) \rangle = 0\,, \\
\langle j (j{-}1\, j{+}1) (i\, i{+}1) (k{-}1\, k) \rangle = 0\,, \\
\langle i\, i {\pm 1}\, j\, k \rangle  = 0\,, \\
\langle i\, j\, j{\pm} 1 \, k\rangle
= 0\,,
\end{split}
\label{eq:subleading}
\end{align}
together with the same collection with $(k{-}1\,k) \to (k\,k{+}1)$,
as well as all
$\langle a\, a{+}1 \, b\, b{+}1 \rangle = 0$ for
$a$, $b$ drawn from the set $\{ i{-}1, i, j{-}1, j, k{-}1, k \}$. 

\paragraph{The maximum codimension boundaries.}
We left this case for last because it is somewhat more subtle.
It is known that the final entries of the symbols of MHV
amplitudes always have the form
$\langle a\, \bar{b} \rangle$~\cite{CaronHuot:2011ky}.
We expect the leading Landau singularity of the maximum codimension
boundary to expose branch points at the vanishing loci of
these final entries.

However,
if we naively solve the Landau equations for the diagram
shown in~\ref{fig3:b},
we run into a puzzle.  The first type of Landau equations~(\ref{eq:LS1})
correspond to the nine cut conditions~(\ref{eq:firstfour})--(\ref{eq:last}),
which of course are satisfied by $\mathcal{L}^{(1)} = (i\,j)$ and
$\mathcal{L}^{(2)} = (i\,k)$.  The second type of Landau
equations~(\ref{eq:LS2}) does not impose any constraints for pentagons
because it is always possible to find a vanishing
linear
combination of the five participating four-vectors.  This naive
Landau analysis
therefore suggests that there is no
leading branch point associated to the
maximum codimension boundary.

This analysis is questionable because, as already noted above,
the Landau diagram associated to the maximal codimension
boundary, shown in figure~(\ref{fig2:b}), 
does not have the form of a valid Feynman diagram.
Therefore it makes little sense to trust the associated Landau analysis.
Instead let us note that the
nine cut conditions~(\ref{eq:firstfour})--(\ref{eq:last})
are not independent; indeed they cannot be as there are only eight
degrees of freedom in the loop momenta.

\begin{figure}[]
\centering
\hspace*{\fill}
\subcaptionbox{\label{newfig:a}}{\includegraphics[scale=1.44]{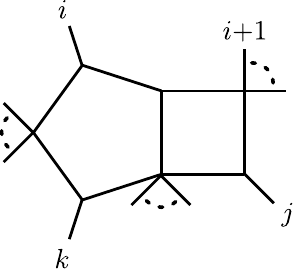}}
\hfill
\subcaptionbox{\label{newfig:b}}{\includegraphics[scale=1.44]{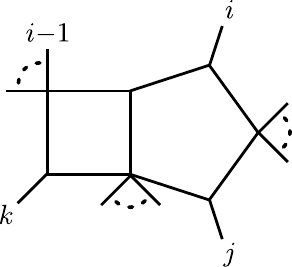}}
\hspace*{\fill}
\caption{Landau diagrams corresponding to all of the cut
conditions~(\ref{eq:firstfour})--(\ref{eq:last}) except for
(a) $\langle \mathcal{L}^{(1)} \, i{-}1 \, i \rangle = 0$,
and
(b) $\langle \mathcal{L}^{(2)} \, i \, i{+}1 \rangle = 0$.
These are the only two cut conditions that are redundant
(each is implied by the other eight, for generic kinematics) and, when omitted,
lead to Landau diagrams that have the form of a standard Feynman integral.
(In both figures $\mathcal{L}^{(1)}$ is the momentum in the right loop and
$\mathcal{L}^{(2)}$ is the momentum in the left loop.)}
\label{newfig}
\end{figure}

We are therefore motivated to identify which of the nine
cut conditions (1) is redundant, in the sense that it is implied by the
other eight for generic external kinematics, and
(2) has the property that when omitted,
the Landau diagram for the remaining eight takes the form of a valid
planar Feynman diagram.
None of the conditions involving $j$ and $k$ shown in
eq.~(\ref{eq:nextfour}) are redundant; all of them must be imposed to
stay on the maximum codimension boundary.  The remaining five
conditions in eqs.~(\ref{eq:firstfour})
and~(\ref{eq:last}) are redundant for general kinematics, but only
two of them satisfy the second property. The corresponding Landau
diagrams are shown in fig.~\ref{newfig}.
Being valid planar Feynman diagrams, the integrand definitely receives
contributions with these topologies (unlike fig.~\ref{fig2:b}), and
will exhibit the associated Landau singularities.

It remains to compute the location of the leading Landau singularities
for these diagrams.  For fig.~\ref{newfig:a}
the on-shell conditions for the pentagon set
$\mathcal{L}^{(2)} = (i\, k)$ while
the Kirkhoff condition for the box is
\begin{equation}
0 = 
\langle j\, (j{-}1\, j{+}1) \mathcal{L}^{(2)} (i\,i{+}1) \rangle=
\langle i\, \bar{j} \rangle \langle i\, i{+}1\, j\, k \rangle\,.
\end{equation}
The Landau equations associated to this topology therefore have solutions
when $\langle i\, \bar{j} \rangle = 0$
or when $\langle i\, i{+}1\, j\, k \rangle = 0$.
However, on the locus $\langle i\, i{+}1\, j\, k \rangle = 0$ it is no longer
true that the eight on-shell conditions shown in fig.~\ref{newfig:a}
imply the ninth condition $\langle \mathcal{L}^{(1)}\, i{-}1\, i\rangle = 0$.
Therefore, this solution of the Landau equations is not relevant to the
maximum codimension boundary.

We conclude that the leading Landau singularity of the maximum codimension
boundary is located on the locus where $\langle i\, \bar{j} \rangle = 0$
or (from fig.~\ref{newfig:b}) $\langle i\, \bar{k} \rangle = 0$.
These results are in agreement with our expectation about the final
symbol entries of MHV amplitudes~\cite{CaronHuot:2011ky}. Relaxations of Figures 7a, 7b at 
$j$, $k$ will not produce any symbol entries.

\paragraph{Conclusion.}

In conclusion, our analysis has revealed that two-loop MHV amplitudes
have physical branch points on the loci of the form
\begin{align} \begin{split}
\langle a \, \bar{b} \rangle = 0\,, \\
\langle a \, b \, c \, c{+}1 \rangle = 0\,, \\
\langle a \, a{+}1\, \bar{b} \cap \bar{c} \rangle = 0\,, \\
\langle a \, (a{-}1 \, a{+}1) (b \, b{+}1) (c \, c{+}1) \rangle = 0\,,
\label{eq:twoloopalphabet}
\end{split} \end{align}
for arbitrary indices $a$, $b$, $c$.
Again let us note that when we say there is a branch point at $x=0$,
we mean a branch cut between $x=0$ and $x=\infty$.
Indeed, this result is in precise accord with the known symbol
alphabet of two-loop MHV amplitudes in SYM theory~\cite{CaronHuot:2011ky}.

\section{Discussion}

In this paper we have improved greatly on the analysis
of~\cite{Dennen:2015bet} by asking the amplituhedron directly to tell
us which branch points of an amplitude are physical.
This analysis requires no detailed knowledge
about how to write formulas for integrands by constructing the canonical
``volume'' form on the amplituhedron.  We only used the amplituhedron's
grossest
feature, which is that it is designed to guarantee that integrands
have no poles outside the space of positive loop configurations.
We have shown in several examples how to use this principle to
completely classify the sets of cut conditions on which integrands
can possibly have support.
Let us emphasize that our proposal
is a completely well-defined geometric algorithm:

\begin{itemize}

\item{Input: a list of the maximal codimension
boundaries of the amplituhedron; for MHV amplitudes
these
are known from~\cite{Arkani-Hamed:2013kca}.}

\item{Step 1: For a given maximal codimension boundary, identify the list of all cut conditions satisfied on this boundary.  For example, at the two-loop boundary shown in figure~\ref{fig3:a}, these would be the nine cut conditions satisfied by the Landau diagram in figure~\ref{fig3:b}, shown in eqs.~(\ref{eq:firstfour})--(\ref{eq:last}). Consider all lower codimension boundaries that can be obtained by relaxing various subsets of these cut conditions, and eliminate those which do not overlap the closure of the amplituhedron, i.e.~those which do not correspond to mutually non-negative configurations of lines $\mathcal{L}^{(\ell)}$.}

\item{Step 2: For each valid set of cut conditions obtained in this manner, solve the corresponding Landau equations~(\ref{eq:LS1}) and~(\ref{eq:LS2}) to determine the location of the corresponding branch point of the amplitude.}

\item{Output: a list of the loci in external kinematic space where the given amplitude has branch points.}

\end{itemize}

As we have mentioned a few times in the text, this algorithm is motivated by
intuition from the Cutkosky rules, according to which
an amplitude's discontinuity is computed by replacing some set of
propagators with delta-functions.  This localizes the integral
onto the intersection of the physical contour and the locus where the
cut conditions are satisfied.
Now is the time to confess that this intuitive motivation is
not a proof of our algorithm,
most notably because the positive kinematic domain lives
in unphysical $(2, 2)$ signature and there is no understanding of how
to make sense of
the physical $i \epsilon$ contour in momentum twistor space
(see however~\cite{Lipstein:2013xra} for work in this direction).
Nevertheless, the prescription works and it warrants serious
further study, in part because it would be very useful to classify the
possible branch points of more general amplitudes in SYM theory.

For amplitudes belonging to the class of generalized polylogarithm
functions (which is believed to contain
at least
all MHV, NMHV and NNMHV amplitudes in SYM theory) the path from
knowledge of branch points to amplitudes is fairly well-trodden.
Such functions can be represented as iterated integrals~\cite{Chen}
and analyzed using the technology of symbols
and coproducts~\cite{Goncharov:2009,Goncharov:2010jf}.
It was emphasized in~\cite{Maldacena:2015iua} that the analytic
structure of an amplitude is directly imprinted on its symbol
alphabet.  In particular, the locus in external kinematic space where
the letters of an amplitude's symbols vanish (or diverge) must exactly
correspond to the locus where solutions of the Landau equations exist.
The above algorithm therefore provides direct information about
the zero locus of an amplitude's symbol alphabet.
For example, the symbol alphabet of one-loop MHV amplitudes must
vanish on the locus~(\ref{eq:oneloopalphabet}), and that
of two-loop amplitudes must vanish on the locus~(\ref{eq:twoloopalphabet}).
Strictly speaking this analysis does not allow one to actually
determine
symbol letters
away from their vanishing locus, but it is encouraging that
in both eqs.~(\ref{eq:oneloopalphabet}) and~(\ref{eq:twoloopalphabet})
the amplituhedron analysis naturally provides the correct symbol
letters on the nose.

In general we expect that only letters of the type
$\langle a\,a{+}1\,b\,b{+}1\rangle$ may appear in the first
entry of the symbol of any amplitude~\cite{Gaiotto:2011dt}.
At one loop, new letters of the type
$\langle a\, \bar{b} \rangle$ begin to appear in the second entry.
At two loops, additional new letters of the type
$\langle a\, (a{-}1\,a{+}1)(b\,b{+}1)(c\,c{+}1)\rangle$
also begin to appear in the second entry, and
new letters of the type  $\langle a\, b\, c\, c{+}1 \rangle$
and $\langle a\, a{+}1 \, \bar{b} \cap \bar{c}\rangle$ begin to appear
in the third.
As discussed at the end of section~3, the final entries
of MHV amplitudes are always
$\langle a\, \bar{b}\rangle$~\cite{CaronHuot:2011ky}.
In our paper we have given almost no thought to the question of where in
the symbol
a given type of letter may begin to appear.
However,
it seems clear that our geometric algorithm can be taken much further
to expose this stratification of branch points, since
the relationship between boundaries of the amplituhedron and Landau
singularities is the same as the relationship between discontinuities
and their branch points.
For example it is clear that at any loop order,
the lowest codimension boundaries
of the amplituhedron that give rise to branch cuts are configurations
where one of the lines $\mathcal{L}$ intersects two lines
$(i\,i{+}1)$ and $(j\,j{+}1)$, with all other lines lying in generic
mutually positive
position.  These configurations give rise
to the expected first symbol entries $\langle i\, i{+}1\, j\, j{+}1\rangle$.
By systematically
following the degeneration of configurations of lines onto
boundaries of higher and higher codimension we expect there should
be a way
to derive the symbol alphabet of an amplitude entry by entry.

In many examples, mere knowledge
of an amplitude's symbol alphabet, together with some other physical
principles, has allowed explicit formulas for the amplitude to be
constructed via a bootstrap approach.
This approach has been particularly powerful for
6-~\cite{Dixon:2011pw,Dixon:2013eka,Dixon:2014voa,Dixon:2014iba,Dixon:2015iva,Caron-Huot:2016owq},
and 7-point~\cite{Drummond:2014ffa}
amplitudes, in which case the symbol alphabet is believed to be
given, to all loop order, by the set of cluster coordinates on the
kinematic configuration space~\cite{Golden:2013xva}.
It would be very interesting to use the algorithm outlined above
to prove this conjecture, or to glean information about symbol
alphabets for more general amplitudes, both MHV and non-MHV.
One simple observation we can make in parting is to note that
although maximum codimension boundaries of the $L$-loop MHV amplitude
involve as many as $2L$ distinct points, the
singularities that arise from genuinely
$L$-loop configurations (rather than products of lower loop order)
involve at most $L+1$ points.
Therefore we
predict that the size of the symbol alphabet of $L$-loop MHV amplitudes
should grow with $n$ no faster than $\mathcal{O}(n^{L+1})$.

It would be very interesting to extend our results to non-MHV amplitudes.
For the N$^K$ amplitude, singularities should still be found only on the boundary of the N$^K$MHV amplituhedron, so the presented approach should still be applicable, albeit more complicated. An important difference would be the existence of poles, in addition to branch points, due to the presence of rational prefactors. We are not certain our approach would naturally distinguish these two types of singularities. However, the singularities of rational prefactors can be found using other means, for example by considering the boundaries of the tree-level amplituhedron.

\acknowledgments

We have benefitted from very stimulating discussions with N.~Arkani-Hamed and are grateful to J. J.~Stankowicz for collaboration on closely related questions and for detailed comments on the draft. MS and AV are grateful to NORDITA and to the CERN theory group for hospitality and support during the course of this work. This work was supported by the US Department of Energy under contract DE-SC0010010 Task A (MS, AV) and Early Career Award DE-FG02-11ER41742 (AV), as well as by Simons Investigator Award \#376208 (AV).

\appendix

\section{Elimination of Bubbles and Triangles}

Here we collect a few comments on the elimination of bubble and
triangle sub-diagrams in the Landau analysis.  These tricks,
together with the results of~\cite{Dennen:2015bet}, can be used
to easily obtain all of the Landau singularities reported
in section~\ref{sec:33}.

\subsection{Bubble sub-diagrams}

The Landau equation for a bubble with propagators $\ell$
and $\ell + p$, which may be a sub-diagram of a larger diagram,
are
\begin{align}
\ell^2 = (\ell + p)^2 &= 0\,, \\
\alpha_1 \ell^\mu + \alpha_2 (\ell + p)^\mu &= 0\,,
\end{align}
where $\alpha_1$ and $\alpha_2$ are the Feynman parameters
associated to the two propagators.  The loop equation has solution
\begin{equation}
\ell^\mu = - \frac{\alpha_2}{\alpha_1 + \alpha_2} p^\mu
\end{equation}
so that
\begin{equation}
\alpha_1 \ell^\mu = - \frac{\alpha_1 \alpha_2}{\alpha_1 + \alpha_2} p^\mu,
\qquad
\alpha_2 (\ell + p)^\mu = \frac{\alpha_1 \alpha_2}{\alpha_1 + \alpha_2} p^\mu,
\end{equation}
while the on-shell conditions simply impose $p^2 = 0$.  Therefore,
we see that any Landau diagram containing this bubble sub-diagram
is equivalent to the same diagram with the bubble replaced by a single
on-shell line with momentum $p^\mu$ and modified Feynman parameter
$\alpha' = \alpha_1 \alpha_2/(\alpha_1+\alpha_2)$.  We do not need to keep
track of the modified Feynman parameter; we simply move on to the
rest of the diagram using the new Feynman parameter $\alpha'$.

In conclusion, any bubble sub-diagram can be collapsed to a single
edge, as far as the Landau analysis is concerned.

\subsection{Triangle sub-diagrams}

Similarly, we will now discuss the various branches associated
to a triangle sub-diagram.  The Landau
equations for a triangle with edges
carrying momenta
$q_1 = \ell$, $q_2 = \ell + p_1 + p_2$ and $q_3 = \ell + p_2$,
and
with corresponding Feynman parameters $\alpha_1$, $\alpha_2$ and $\alpha_3$,
are
\begin{align}
\ell^2 = (\ell + p_2)^2 = (\ell + p_1 + p_2)^2 &= 0\,,
\label{eq:onshell}
\\
\alpha_1 \ell^\mu + \alpha_2 (\ell + p_1 + p_2)^\mu + \alpha_3
(\ell + p_2)^\mu &= 0\,.
\end{align}
The solution to the loop equation is
\begin{equation}
\ell^\mu = - \frac{(\alpha_2 + \alpha_3) p^\mu_2 + \alpha_2 p^\mu_1}{\alpha_1 + \alpha_2 + \alpha_3}
\end{equation}
while eqs.~(\ref{eq:onshell}) impose the two conditions
\begin{align}
0 &= p_1^2\, p_2^2\, p_3^2\,, \\
(\alpha_1 : \alpha_2 : \alpha_3) &=
\left( p_1^2 (-p_1^2 + p_2^2 + p_3^2) :
p_2^2 (p_1^2 - p_2^2 + p_3^2) :
p_3^2 (p_1^2 + p_2^2 - p_3^2) \right)\,
\end{align}
where $p_3 = - p_1 - p_2$.  Suppose we follow the branch
$p_1^2 = 0$.  In this case $\alpha_1$ is forced to vanish, effectively
reducing the triangle to a bubble with edges
\begin{equation}
\alpha_2 q_2^\mu = \frac{\alpha_3 p_2^2}{p_2^2 - p_3^2} p_1^\mu\,,\qquad
\alpha_3 q_3^\mu = - \frac{\alpha_3 p_2^2}{p_2^2 - p_3^2} p_1^\mu\,.
\end{equation}
This is equivalent (by appendix A.1)
to a single on-shell line carrying momentum
$p_1^\mu$.
A similar conclusion clearly holds for the branches
$p_2^2 = 0$ or $p_3^2 = 0$.  If any two of $p_1^2$, $p_2^2$
or $p_3^2$ simultaneously vanish, then the two corresponding
Feynman parameters must vanish.
Finally, if all three $p_i^2$ vanish, then the Landau equations are
identically satisfied for any values of the three $\alpha_i$.
In conclusion,
triangle sub-diagrams of a general Landau diagram can
be analyzed by considering separately each of the seven branches
outlined here.

\end{document}